\documentclass[10pt,conference]{IEEEtran}
\IEEEoverridecommandlockouts
\usepackage{cite}
\usepackage{amsmath,amssymb,amsfonts}
\usepackage{graphicx}
\usepackage{textcomp}
\usepackage{xcolor}
\usepackage{array}
\usepackage{comment}

\usepackage{siunitx}
\usepackage{multirow}

\usepackage[ruled]{algorithm2e}
\usepackage{algpseudocode}

\def\BibTeX{{\rm B\kern-.05em{\sc i\kern-.025em b}\kern-.08em
    T\kern-.1667em\lower.7ex\hbox{E}\kern-.125emX}}

\begin{document}

\title{Efficient Quantum Circuit Encoding of Object Information in 2D Ray Casting

\thanks{This work was supported in part by the National Research Foundation of Korea (NRF) funded by the Korean government (Ministry of Science and ICT) under Grant RS-2024-00358298, and in part by Grant RS-2024-00407003 from the ``Development of Advanced Technology for Terrestrial Radionavigation System'' project, funded by the Ministry of Oceans and Fisheries, Republic of Korea.} 
} 

\author{\IEEEauthorblockN{Seungjae Lee} 
\IEEEauthorblockA{\textit{School of Integrated Technology} \\
\textit{Yonsei University}\\
Incheon, Korea \\
seungjae@yonsei.ac.kr} 
\and
\IEEEauthorblockN{Suhui Jeong} 
\IEEEauthorblockA{\textit{School of Integrated Technology} \\
\textit{Yonsei University}\\
Incheon, Korea \\
ssuhui@yonsei.ac.kr} 
\and
\IEEEauthorblockN{Jiwon Seo${}^{*}$} 
\IEEEauthorblockA{\textit{School of Integrated Technology} \\
\textit{Yonsei University}\\
Incheon, Korea \\
jiwon.seo@yonsei.ac.kr}
{\small${}^{*}$ Corresponding author}
}

\maketitle

\begin{abstract}
Quantum computing holds the potential to solve problems that are practically unsolvable by classical computers due to its ability to significantly reduce time complexity.
We aim to harness this potential to enhance ray casting, a pivotal technique in computer graphics for simplifying the rendering of 3D objects.
To perform ray casting in a quantum computer, we need to encode the defining parameters of primitives into qubits.
However, during the current noisy intermediate-scale quantum (NISQ) era, challenges arise from the limited number of qubits and the impact of noise when executing multiple gates.
Through logic optimization, we reduced the depth of quantum circuits as well as the number of gates and qubits.
As a result, the event count of correct measurements from an IBM quantum computer significantly exceeded that of incorrect measurements.
\end{abstract}

\begin{IEEEkeywords}
 Quantum computing, ray casting, logic optimization
\end{IEEEkeywords}

\section{Introduction}

Since Richard Feynman's proposal of quantum computing, the field has seen significant advancements \cite{Feynman1982:Simulating}.
Developments within the theoretical framework of quantum computing include Deutsch's quantum Turing machine \cite{Deutsch1985:Quantum}, Shor's factorization algorithm \cite{Shor1997:Polynomial}, and the quantum Fourier transform algorithm \cite{Coppersmith2002:Approximate}, among others \cite{Kitaev1995:Quantum, Bennett1997:Strengths}.
On the technological front, several companies have built actual quantum computers. 
However, as the term noisy intermediate-scale quantum (NISQ) suggests, the current state of quantum computing technology faces issues such as limited coherence time, an insufficient number of qubits, and significant error rates in gates \cite{Preskill2018:Quantum, Michael2019:Beyond}.

Interest in quantum computing remains strong due to the significant time-saving benefits for some problems provided by quantum parallelism and the superposition of states \cite{Marinescu2005:Promise, Bennett1997:Strengths}.
For certain problems, algorithms that leverage these quantum characteristics have been proposed; for example, Grover's algorithm is specifically designed for unstructured search problems \cite{Grover1996:A_fast_quantum}.

Studies have applied Grover's algorithm to ray casting \cite{Alves2019:Quantum, Haines2019:Ray, Glassner2001:Andrew}. 
Ray casting is a widely used technique for rendering 3D graphics onto a 2D screen. 
It works by calculating which of the primitives intersects first with the ray corresponding to each pixel on the display, thereby deciding what to output for that pixel.
Given a ray corresponding to a single pixel on the display, the ray casting process checks whether the ray intersects with each of the primitives.
In this context, primitives refer to simple geometric shapes used in computer graphics, such as polygons, spheres, and rectangles.

Many of the graphic techniques widely used today, such as ray tracing \cite{Kim23:Machine, Kim22:Machine, Lee22:Urban, Kim23:Single} for satellite navigation \cite{Lee23:Seamless, Lee22:Optimal, Sun21:Markov, Kim21:GPS, Jia21:Ground, Park21:Single, Lee24:A}, are applications of ray casting, making it very important. 
However, it has always been limited by its computational complexity, as ray casting requires checking the intersection of every ray with all predefined primitives. 
Thus, there is interest in improving the process of intersection verification for ray casting by utilizing quantum parallelism.

For instance, the concept of ray/sphere intersection verification using a quantum computer has been proposed under the assumption that a quantum computer with sufficient qubits is available \cite{Glassner2001:Andrew, Haines2019:Ray}.
Alves \textit{et al.} \cite{Alves2019:Quantum} and Santos \textit{et al.} \cite{Santos2022:Towards} proposed quantum algorithms for ray casting in a simplified scenario where all primitives are axis-aligned rectangles, due to the limitation in the number of qubits of currently available quantum computers.
Axis-aligned rectangles refer to rectangles whose sides are parallel to the coordinate axes.
Specifically, Alves \textit{et al.} \cite{Alves2019:Quantum} presented an algorithm for the case of an orthographic camera, while Santos \textit{et al.} \cite{Santos2022:Towards} extended this algorithm to the case of a pinhole camera, which requires the representation of floating point numbers due to the presence of a 3D direction vector, and further proposed methods to reduce errors.

The studies above \cite{Alves2019:Quantum, Santos2022:Towards}, however, did not specify a generalized optimization technique to reduce the number of gates when initializing the parameters of the primitives, such as the positions of the four sides of the rectangles, in a quantum circuit.
These algorithms require assigning an integer, referred to as an index $(0, 1, \ldots, N-1)$, to each of the given $N$ primitives, and implementing a function that takes an index as input and outputs numbers, known as parameters, describing the corresponding primitive.
In the scenario considered in \cite{Alves2019:Quantum, Santos2022:Towards}, since all the primitives are axis-aligned rectangles, each rectangle can be specified by just the $x$-coordinates of its left and right sides, and the $y$-coordinates of its bottom and top sides, making these the rectangle's parameters.

We focused on the single solution case of \cite{Alves2019:Quantum}, where a ray intersects with only one primitive.
In our case, all rays are set to be parallel to the $z$-axis, and thus the $z$ value of each primitive represents the distance between the ray’s starting point and the primitive.
However, since it is a single solution case, there is no need to determine which primitive is closest to the starting point of the ray; it is only necessary to verify whether the ray intersects with a primitive.
Therefore, the $z$ value of each primitive is not used as a parameter.

In this study, we applied a generalized optimization technique to minimize the number of gates necessary for a quantum circuit and verified that our implementation of a quantum circuit, which takes each index from the list of primitives as input and outputs its parameters, worked correctly.
Additionally, we confirmed that the applied optimization technique significantly increases the probability of the circuit operating correctly on a real NISQ quantum computer compared to cases without optimization.
All experiments in this study were conducted on IBM's quantum computing framework ``Qiskit'' and its compute resource ``ibm\_torino.''

\section{Geometric Setup}
\label{sec:Setup}

\begin{figure}
    \centering
    \includegraphics[width=0.95\linewidth]{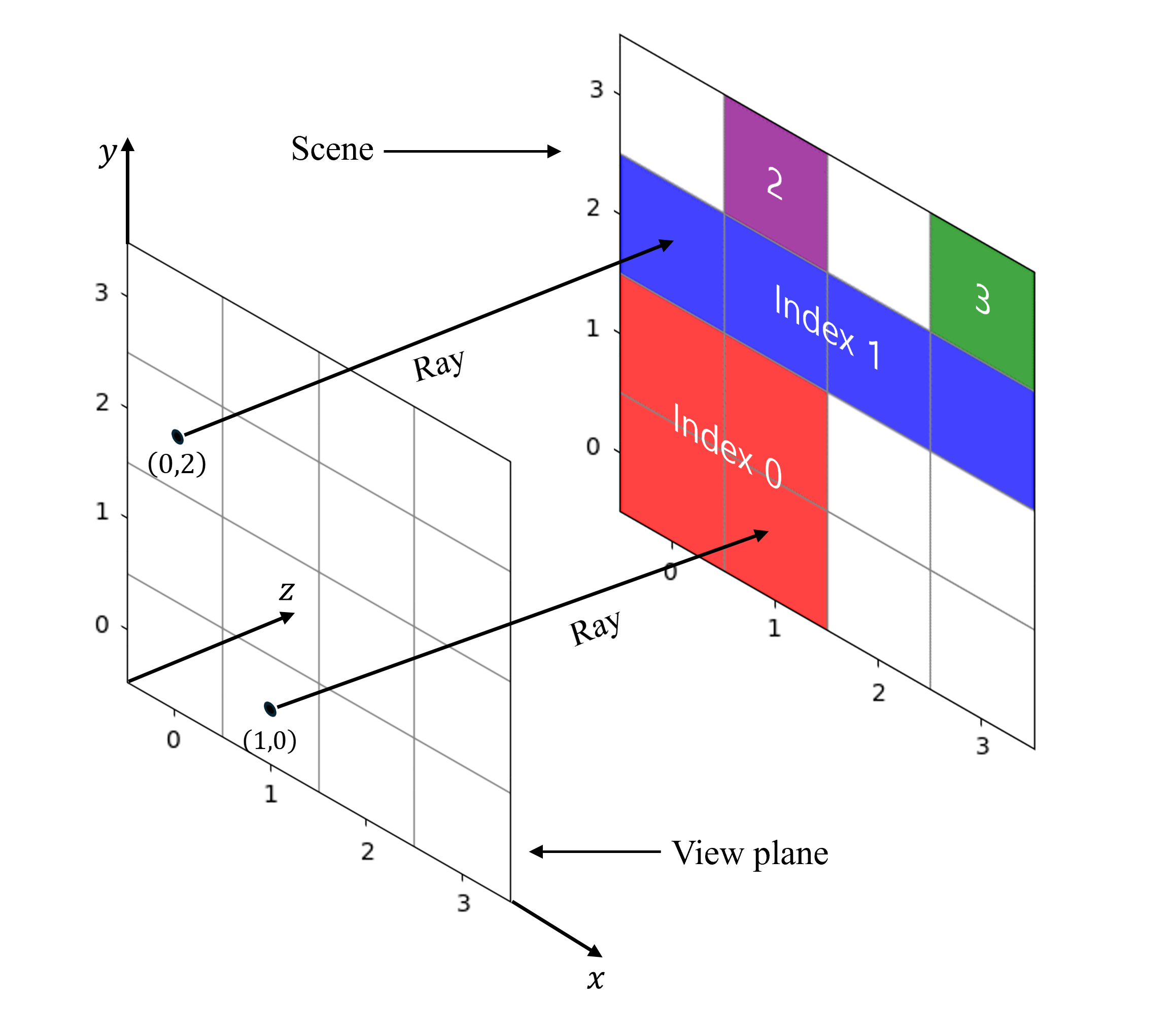}
    \caption{Geometric configuration with $b_{x}=b_{y}=4$ and four primitives (i.e., axis-aligned rectangles in red, blue, purple, green) parameterized as $(0,1,0,1)$, $(0,3,2,2)$, $(1,1,3,3)$, $(3,3,3,3)$, respectively. This setup follows the configuration of \cite{Alves2019:Quantum}.}
    \label{fig:444_config}
\end{figure}

A scene for primitives can be defined using a set
\begin{equation} 
    \{(x,y)\in\mathbb{Z}^{2}\mid0\leq x<b_{x},0\leq y<b_{y}\}
\end{equation}
where $b_{x}$ and $b_{y}$ are fixed integers representing the bounds of $x$ and $y$, respectively.
As previously mentioned, the $z$ value is irrelevant because we are dealing with the single solution case.
The view plane, used to project what is visible, is contained within the plane of $z=0$, and all primitives in this study are axis-aligned rectangles that are parallel to the view plane.
Rays are lines that are perpendicular to the view plane at $z=0$, and both the $x$ and $y$ coordinates of the starting points of rays are integers.
Since primitives considered in this study are all axis-aligned rectangles, each primitive can be represented as a set
\begin{equation}
\{(x,y)\mid m_{x}\leq x\leq M_{x},m_{y}\leq y\leq M_{y}\}
\end{equation}
where $m_x$ and $M_x$ denote the $x$-coordinates of the left and right sides of the rectangle, respectively, and $m_y$ and $M_y$ represent the $y$-coordinates of the bottom and top sides of the rectangle, respectively.

Fig. \ref{fig:444_config} illustrates an example of the setup, where the bounds $b_x$ and $b_y$ for the $x$ and $y$ coordinates of the scene are both 4, and the zeroth, first, second, and third primitives are parameterized as $(m_x,M_x,m_y,M_y)=(0,1,0,1), (0,3,2,2), \\(1,1,3,3)$, and $(3,3,3,3)$ and colored in red, blue, purple, and green, respectively.
We can see in Fig. \ref{fig:444_config} that a ray originating from $(x,y)=(1,0)$ intersects with the zeroth primitive and a ray starting from $(0,2)$ intersects with the first primitive.

\section{Quantum Circuit Implementation}
\label{sec:Implement}

To implement a quantum circuit that corresponds to a function accepting a primitive index as input and outputs the parameters of the respective primitive, it is necessary to allocate qubits for both the index and each parameter associated with the primitive.
For example, when the circuit takes index $0$ in Fig. \ref{fig:444_config} as its input, the output of the circuit should be parameters $(0,1,0,1)$ that correspond to the zeroth primitive. 
The quantum circuit needs registers comprising ${\log_{2}{N}}$ qubits where $N$ is the number of primitives (e.g., $N=4$ in Fig. \ref{fig:444_config}), ${\log_{2}{b_{x}}}$ qubits to represent each $m_{x}$ or $M_{x}$ value of a primitive, and ${\log_{2}{b_{y}}}$ qubits to represent each $m_{y}$ or $M_{y}$ value of a primitive (e.g., $b_x = 4, b_y = 4$ in Fig. \ref{fig:444_config}), along with additional auxiliary qubits for implementation.

To illustrate the process of assigning values to a designated register for $m_{x}$, consider the example in Fig. \ref{fig:444_config}, where the values of $m_{x}$ of four primitives are 0, 0, 1, and 3.
It is necessary to implement a quantum circuit representing a function that maps the index of a primitive to the $m_{x}$ value of the primitive, which is $f=(f_{2},f_{1} ):\{0, 1\}^{2}\rightarrow\{0, 1\}^{2}$, defined as follows:
\begin{equation}
\label{eq:f}
\begin{aligned}
        &(0,0)\mapsto(0,0),\quad (0,1)\mapsto(0,0) \\
        &(1,0)\mapsto(0,1),\quad (1,1)\mapsto(1,1).
\end{aligned}
\end{equation}
The input and output of the function $f$ are represented in binary numbers.
Here, $f_1$ represents the least significant bit, and $f_2$ represents the most significant bit.
For example, when the input is $(1,0)$, which is $1 \times 2^1 + 0 \times 2^0 = 2$ (i.e., input is index $2$, representing the second primitive), the output should be $(0,1)$, which is $0 \times 2^1 + 1 \times 2^0 = 1$ (i.e., output is $m_{x}=1$, corresponding to the second primitive).

From this, it is possible to derive a sum of products (SOP) for each component, $f_1$ or $f_2$, of $f$.
In the example above, $f_1$ outputs $1$ when the input $x=(x_2,x_1)$ is either $x=(1,0)$ or $x=(1,1)$. 
Therefore, for $f_1$ to be true, it must be that $x_2$ is true and $x_1$ is false, or $x_2$ is true and $x_1$ is true. 
Thus, $f_1$ can be expressed as $x_2x_1'+x_2x_1$, where $x_i$ signifies the value being true, and $x_i'$ represents the value being false.

Since we have obtained a logical expression consisting of AND, OR, and NOT operations, the corresponding gates can be applied in a quantum circuit. 
It is well-known that the NOT gate corresponds to the Pauli $X$ gate in quantum circuits, which does not require an additional qubit. 
However, the situation differs for AND or OR gates. 
For the AND operation, if the target qubit is in the initial state $\lvert 0\rangle$ before the gate operation, it can be realized using the Toffoli gate (also known as the CCX gate), which acts as an AND operator with two control qubits producing a target output \cite{Nielsen2010:Quantum}. 
Similarly, the OR operation can be achieved by applying gates in the sequence of $(X\otimes X\otimes I), CCX, (X\otimes X\otimes X)$, where the last qubit serves as the target in this circuit configuration. 
Each time the quantum circuit employs an AND or OR operation, the number of qubits needed for the circuit increases by one. 
Quantum computing is functionally complete with respect to Boolean arithmetic, enabling it to execute any computation expressible on classical computers, provided there are sufficiently many qubits \cite{Shaik20:Implementation}.

\section{Logic Optimization}
\label{sec:Logic}

In Section \ref{sec:Implement}, we demonstrated the feasibility of implementing the quantum circuit for the function $f$ in (\ref{eq:f}) by deriving the corresponding SOPs. 
However, implementing the function using the SOPs directly is not recommended.
We have to use auxiliary qubits to store the results of each product in the SOPs.
Subsequently, the results of applying OR operations among those auxiliary qubits are used to represent the final results of each component function for each parameter.
However, in the SOPs we derived, when the number of qubits representing the index of primitives is denoted as $n$, it can consist of up to $2^n$ products in the worst-case scenario. This means that an exponential number of auxiliary qubits may be needed.

Additionally, in actual quantum computers, each gate must be decomposed into the elementary gates supported by the respective computer\cite{Barenco1995:Elementary}.
For example, ``ibm\_torino'' decomposes gates into controlled-Z ($CZ$), identity ($I$), $R_z$, $SX$, and $X$ gates.
Here, $R_z$ is a gate that performs rotation around the $z$-axis on the Bloch sphere, and $SX$ is the square root of the $X$ gate, which becomes the $X$ gate when applied twice.
When implementing the SOPs, each product composed of $n$ terms requires a multi-controlled $X$ gate with $n$ controls.
Decomposing this into elementary gates results in the use of significantly more elementary gates as $n$ increases.
Therefore, it is necessary to perform logic optimization to find an SOP that maintains the same expression while reducing both the number of terms in each product and the number of products.

By performing logic optimization, the number of products constituting each SOP and the number of terms composing each product can be reduced.
Reducing the number of products decreases the number of auxiliary qubits used, and reducing the number of terms decreases the number of elementary gates used, which is beneficial given the current era of NISQ quantum computers characterized by a limited number of qubits and relatively high gate error rates.
We conducted this logic optimization using the Quine-McCluskey (QMC) algorithm\cite{McCluskey1956:Minimization} and Petrick's method\cite{Petrick1956:Direct}.
Their implementations are relatively simple and guarantee optimal results.

To address the space complexity on quantum computers, it is important to perform such logic optimizations without significantly compromising the temporal advantages offered by quantum computers.
When the number of inputs to the function to be optimized is $n$, it is known that the number of outputs of QMC is $\Omega(3^n/n)$ and $O(3^n/\sqrt{n})$\cite{Chandra1978:Number}.
Therefore, for large-scale problems, the use of suboptimal heuristic logic minimizers such as ESPRESSO\cite{Rudell1987:Multiple} is recommended.

The QMC algorithm optimizes the implementation of a given Boolean function $g:\{0,1\}^n\rightarrow\{0,1\}$ by combining minterms into prime implicants.
A minterm consisting of $n$ variables $x_{n-1}, \ldots, x_0$ refers to a form obtained by taking each $x_i$ either as itself or its complement $x_i'$ and then performing a logical AND on all of them.
Therefore, when considering each minterm as a Boolean function, the combination of the $n$ bits that makes it true is unique.

The Boolean function $g$ can now be expressed as the logical disjunction of the minterms corresponding to each input that results in an output of $1$.
For example, assume $n=3$ and the necessary and sufficient condition for $g(x)=1$ is that $x$ corresponds to the binary representations of $1,3,7$.
Then, $g=x_2'x_1'x_0 + x_2'x_1x_0 + x_2x_1x_0$, where the prime symbol ($'$) indicates the complement of the variable.

Now, let $m(i)$ denote the minterm for the binary representation of $i=0,1,\ldots,2^n-1$.
By comparing the minterms used to express $g$, if they differ by only one variable being complemented, they can be combined into a single expression.
In the above example, $x_2'x_1'x_0$ and $x_2'x_1x_0$ differ in $x_1$ in this manner, so they can be combined as $x_2'x_0=x_2'x_1'x_0+x_2'x_1x_0$.

A product combined in this way is called an implicant and is denoted by $m(\ldots)$, where the parentheses indicate the implicant is true for the binary representations of the numbers within the parentheses.
The number of values an implicant can represent is called its size, and a minterm can be considered an implicant of size $1$.
Consequently, the function $g$ can now be expressed as $g=m(1,3)+m(3,7)$.

The QMC algorithm continues this process by comparing implicants of the same size to create larger implicants.
This process is repeated until no new implicants can be formed, identifying the implicants that can no longer combine with others as prime implicants.

Now, what remains is the process of selecting some of the prime implicants.
Suppose the function $h:\{0,1\}^3\rightarrow\{0,1\}$ outputs $1$ for the binary representations of $0, 1, 2, 5, 6$, and $7$.
When optimizing this function using the QMC algorithm, the prime implicants obtained are $m(0,1)$, $m(0,2)$, $m(1,5)$, $m(2,6)$, $m(5,7)$, and $m(6,7)$.
In other words, $h=m(0,1)+m(0,2)+m(1,5)+m(2,6)+m(5,7)+m(6,7)$.
The fact that selecting and summing $m(0,1)$, $m(2,6)$, and $m(5,7)$ from these prime implicants, in the form $m(0,1) + m(2,6) + m(5,7)$, is still equal to $h$ clearly shows that the process of selecting among the prime implicants remains.

Petrick’s method is one of the strategies for enhancing solution optimality.
Petrick's method minimizes the logical expressions by constructing a logical expression that is always false using prime implicants.
Here is an example of a logical expression that is false for every number.
Consider a logical expression formed by the conjunction of $m(0,1)$ and $m(2,6)$.
Here, $m(0,1) = x_2'x_1'$ and $m(2,6) = x_1x_0'$.
The conjunction of $m(0,1)$ and $m(2,6)$ would be true for any number that satisfies both $x_2'x_1'$ and $x_1x_0'$.
However, there is no number whose binary representation has $x_1$ being both 0 and 1.
Therefore, this logical expression is false for all numbers.

Petrick’s method combines prime implicants to create a product of sums (POS) that is always false and then expands this to form a SOP.
The algorithm selects the smallest product from this SOP.
Specifically, for each number whose binary representation yields an output of $1$ in the function $g$ we aim to implement, we construct a sum by combining all prime implicants that produce an output of $1$ when given the binary representation of the number as input.
For example, in the case of implementing the function $h$ mentioned earlier, the prime implicants that yield a value of $1$ when the binary representation of $0$ is input are $m(0,1)$ and $m(0,2)$, and the prime implicants when the binary representation of $1$ is input are $m(0,1)$ and $m(1,5)$.
Therefore, we construct $m(0,1)+m(0,2)$ and $m(0,1)+m(1,5)$, respectively.
Subsequently, AND operations are performed among all these sums to create a product of sums.
In the case of the function $h$, the final expression of the POS we aimed to construct is as follows:
\begin{equation}
\label{eq:expr}
\begin{aligned}
    (m(0,1)+m(0,2))\cdot(m(0,1)+m(1,5))\cdot\\
    (m(0,2)+m(2,6))\cdot(m(1,5)+m(5,7))\cdot\\
    (m(2,6)+m(6,7))\cdot(m(5,7)+m(6,7))
\end{aligned}
\end{equation}
This combination results in a term that is true only if there exists a prime implicant capable of representing all numbers.
This occurs when the QMC algorithm finds a single prime implicant that represents all and only the numbers we are looking for.
In such cases, the optimal solution means a sum composed of the smallest number of prime implicants that still equals the desired function, so it is evident that QMC alone has found the optimal solution, and there is no need to use Petrick’s method.
In all other cases, the above POS evaluates to 0.
Utilizing De Morgan's laws to expand this formula into a SOP and then simplifying it with the identities $A+AB=A$, $AA=A$, and $A+A=A$ where $A,B$ are Boolean variables, the expanded expression remains false, indicating that each product in the expression is false.

While each product in the SOP is false in itself, a term resulting from replacing all AND operations within any product in the SOP with OR operations can represent all the numbers we are looking for.
Therefore, choosing a product of the shortest length offers an optimal solution in logic optimization.

\section{Results} 

\subsection{Results from Simulator}

\begin{figure}
    \centering
    \includegraphics[width=0.8\linewidth]{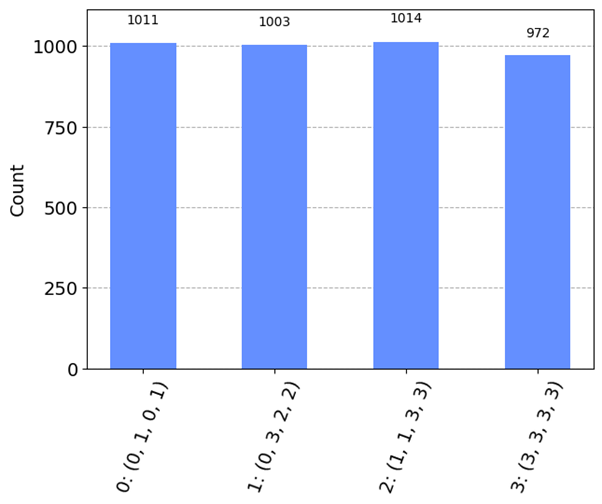}
    \caption{Measurement results of the optimized quantum circuit for the configuration with $b_x=b_y=4$, as obtained from the simulator. The histogram indicates the event count of each input (i.e., primitive index $i$) and output (i.e., parameters) pair, denoted as $i : (m_x,M_x,m_y,M_y)$.}
    \label{fig:444_simul}
\end{figure}

\begin{table}
\centering
\caption{Depth, number of gates, and number of qubits in the configuration with $b_x=b_y=4$ (simulator case)}
\label{tab:444_table}
\begin{tabular}{|l|c|c|} 
\hline
& Optimized & Not optimized \\ \hline
Depth & 41 & 68 \\ \hline
\# Gates & 64 & 126 \\ \hline
\# Qubits & 12 & 14 \\ \hline
\end{tabular}
\end{table}

First, we confirmed that the quantum circuits we implemented operate properly on the ``ibmq\_qasm\_simulator,'' a cloud quantum simulator provided by IBM.
This was conducted to verify the performance of the circuit in an ideal environment free from any noise and interference.
We analyzed the results in the two specific configurations used by \cite{Alves2019:Quantum}: one with $b_x=b_y=4$ and four primitives parameterized as $(0,1,0,1)$, $(0,3,2,2)$, $(1,1,3,3)$, $(3,3,3,3)$, and the other with $b_x=b_y=8$ and eight primitives parameterized as $(1,3,1,2)$, $(6,6,1,4)$, $(0,3,7,7)$, $(7,7,0,0)$, $(1,2,4,5)$, $(4,4,0,2)$, $(4,4,4,7)$, $(7,7,5,7)$.
In both configurations, the primitives were parameterized as the sequence of $(m_{x},M_{x},m_{y},M_{y})$ values.

Fig. \ref{fig:444_simul} presents the outcome of measuring the first configuration with four primitives with logic optimization. 
The quantum circuit was set up so that the probability of measuring each primitive index (i.e., 0, 1, 2, 3 for each of the four primitives) is uniformly distributed in a single measurement.
To validate the correct operation of the quantum circuit, we conducted 4000 measurements. 
The histogram illustrates the event count of each input (i.e., primitive index $i$) and output (i.e., parameters) pair, denoted as $i: (m_x,M_x,m_y,M_y)$. 
For example, the event count for measuring the input $i=0$ and the output $(m_x,M_x,m_y,M_y)=(0,1,0,1)$ was 1011.

As mentioned earlier, since the four primitives were parameterized as $(0,1,0,1)$, $(0,3,2,2)$, $(1,1,3,3)$, $(3,3,3,3)$, the correct output for the primitive index $i=0$ is $(0,1,0,1)$. 
Similarly, $i: (m_x,M_x,m_y,M_y) = 1: (0,3,2,2)$ in Fig. \ref{fig:444_simul} also represents a correct input and output pair. 
The remaining two pairs, $2: (1,1,3,3)$ and $3: (3,3,3,3)$, are also correct pairs. 
The event counts of each correct measurement were 1011, 1003, 1014, and 972, respectively, and their sum is equal to the total number of measurements, 4000. 
Therefore, it is evident that there were no incorrect measurements in the 4000 measurements, which confirms the correct implementation of the quantum circuit.

When comparing the performance of the circuits, it is essential to consider factors such as the depth of the circuits, the number of gates applied, and the number of qubits used in each case. 
These metrics are summarized in Table \ref{tab:444_table}. 
The depth decreased from 68 to 41, the number of gates was reduced from 126 to 64, and the number of qubits decreased from 14 to 12 when the circuit was optimized using the method outlined in Section \ref{sec:Logic}.

Since $\log_{2}{N}$ qubits are needed to represent an index, $\log_{2}{b_x}$ qubits for $m_x$ and $M_x$, and $\log_{2}{b_y}$ qubits for $m_y$ and $M_y$, a total of $\log_{2}{N} + \log_{2}{b_x} + \log_{2}{b_x} + \log_{2}{b_y} + \log_{2}{b_y} = 5 \log_{2}{4} = 10$ qubits are used in the quantum circuit for the first configuration.
As summarized in Table \ref{tab:444_table}, the total number of qubits used in the optimized case was 12, and we understand that 10 qubits are needed to represent the index and parameters. 
Thus, the number of auxiliary qubits for this case was 2. 
After the logic optimization, the number of auxiliary qubits decreased from 4 to 2.

\begin{table}
\centering
\caption{Depth, number of gates, and number of qubits in the configuration with $b_x=b_y=8$ (simulator case)}
\label{tab:888_table}
\begin{tabular}{|l|c|c|} 
\hline
& Optimized & Not optimized \\ \hline
Depth & 93 & 157 \\ \hline
\# Gates & 201 & 363 \\ \hline
\# Qubits & 19 & 21 \\ \hline
\end{tabular}
\end{table}

Table \ref{tab:888_table} summarizes the depth of the circuits, the number of gates applied, and the number of qubits used for the second configuration with eight primitives when the logic optimization was applied and not applied.

\subsection{Results from Quantum Computer}

\begin{figure}
    \centering
    \includegraphics[width=0.9\linewidth]{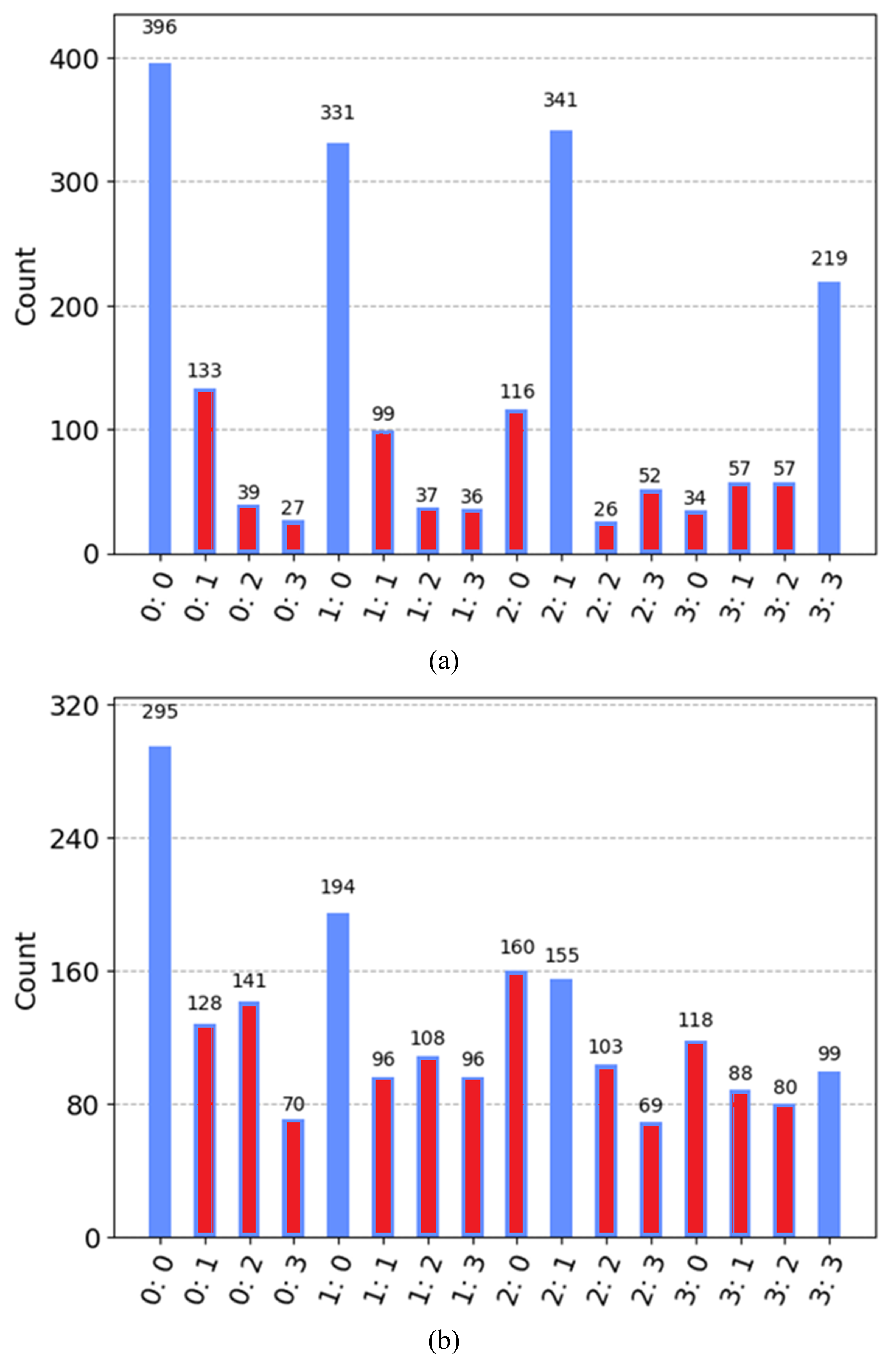}
    \caption{Measurement results from a quantum computer when (a) the logic optimization was applied and (b) not applied. 
    The histogram indicates the event count of each input (i.e., primitive index $i$) and output (i.e., parameter $m_x$) pair, denoted as $i : m_x$.
    Incorrect results are indicated by red color.}
    \label{fig:real}
\end{figure}

\begin{table}
\centering
\caption{Depth, number of gates, and number of qubits in the configuration with $b_x=b_y=4$ (quantum computer case)}
\label{tab:real}
\begin{tabular}{|l|c|c|} 
\hline
& Optimized & Not Optimized \\ \hline
Depth & 62 & 680 \\ \hline
\# Gates & 107 & 1245 \\ \hline
\# Qubits & 7 & 10 \\ \hline
\end{tabular}
\end{table}

In actual quantum computers, conducting measurements encounters significant differences from theoretical simulations due to various interferences.
Attempting to measure all parameters as in simulations would lead to a large number of measurement scenarios, $b_{x}^{2} b_{y}^{2} N$, due to errors.
For example, while parameters $(1,3,1,2)$ are the correct output for index $0$, the actual output from a quantum computer can range anywhere between $(0,0,0,0)$ and $(7,7,7,7)$. 
Furthermore, decomposing various required gates into elementary gates of the quantum computer is necessary.
This conversion increases the circuit's depth and the number of applied gates compared to the simulator circuit.

We implemented a quantum circuit in the configuration where $b_x=b_y=4$, which outputs only two parameters, $m_x$ and $M_x$.
We focused solely on observing the results for $m_x$.
This strategy aimed to enhance accuracy in noisy conditions by reducing the number of gates, entanglements, and the circuit’s depth.

Fig. \ref{fig:real} illustrates the measurement outcomes from an IBM quantum computer.
Fig. \ref{fig:real}(a) depicts the optimized scenario, while Fig. \ref{fig:real}(b) shows the non-optimized one.
Both were set to a measurement count of 2000.
The optimized scenario prominently featured the correct values, with other values occurring significantly less often. 
In contrast, the non-optimized one displayed seriously distorted results.

Table \ref{tab:real} presents each circuit’s depth, qubit usage, and gate usage in the quantum computer, which demonstrates notable reductions in all aspects through optimization. 
The number of elementary gates used during the decomposition of multi-controlled gates increases with the number of controls. 
Before optimization, the multi-controlled gates had many controls, leading to the use of a large number of elementary gates. 
However, after optimization, the number of controls for multi-controlled gates is significantly reduced.

\section{Conclusion}

In this paper, we optimized the process of representing the parameters of primitives for a simplified 2D ray casting scenario in a quantum computer and evaluated its impact on the measurement results. 
Through such logic optimization, we observed a substantial reduction in the depth of quantum circuits, as well as the number of gates and qubits required in actual quantum computers. 
These reductions were found to have a significant impact on the measurement results.

\bibliographystyle{IEEEtran}
\bibliography{IUS_publications, mybibfile, JPNT}

\begin{thebibliography}{10}
\providecommand{\url}[1]{#1}
\csname url@samestyle\endcsname
\providecommand{\newblock}{\relax}
\providecommand{\bibinfo}[2]{#2}
\providecommand{\BIBentrySTDinterwordspacing}{\spaceskip=0pt\relax}
\providecommand{\BIBentryALTinterwordstretchfactor}{4}
\providecommand{\BIBentryALTinterwordspacing}{\spaceskip=\fontdimen2\font plus
\BIBentryALTinterwordstretchfactor\fontdimen3\font minus \fontdimen4\font\relax}
\providecommand{\BIBforeignlanguage}[2]{{%
\expandafter\ifx\csname l@#1\endcsname\relax
\typeout{** WARNING: IEEEtran.bst: No hyphenation pattern has been}%
\typeout{** loaded for the language `#1'. Using the pattern for}%
\typeout{** the default language instead.}%
\else
\language=\csname l@#1\endcsname
\fi
#2}}
\providecommand{\BIBdecl}{\relax}
\BIBdecl

\bibitem{Feynman1982:Simulating}
R.~P. Feynman, ``Simulating physics with computers,'' \emph{Feynman and Computation}, pp. 133--153, 1982.

\bibitem{Deutsch1985:Quantum}
D.~Deutsch, ``Quantum theory, the {Church–Turing} principle and the universal quantum computer,'' \emph{Proc. R. Soc. Lond. A}, vol. 400, pp. 97--117, 1985.

\bibitem{Shor1997:Polynomial}
P.~W. Shor, ``Polynomial-time algorithms for prime factorization and discrete logarithms on a quantum computer,'' \emph{SIAM J. Comput.}, vol.~26, no.~5, pp. 1484--1509, 1997.

\bibitem{Coppersmith2002:Approximate}
D.~Coppersmith, ``An approximate {Fourier} transform useful in quantum factoring,'' arXiv:quant-ph/0201067, 2002.

\bibitem{Kitaev1995:Quantum}
A.~Y. Kitaev, ``Quantum measurements and the {Abelian} stabilizer problem,'' arXiv:quant-ph/9511026, 1995.

\bibitem{Bennett1997:Strengths}
C.~H. Bennett, E.~Bernstein, G.~Brassard, and U.~Vazirani, ``Strengths and weaknesses of quantum computing,'' \emph{SIAM J. Comput.}, vol.~26, no.~5, pp. 1510--1523, 1997.

\bibitem{Preskill2018:Quantum}
J.~Preskill, ``Quantum computing in the {NISQ} era and beyond,'' \emph{Quantum}, vol.~2, p.~79, 2018.

\bibitem{Michael2019:Beyond}
M.~Brooks, ``Beyond quantum supremacy: The hunt for useful quantum computers,'' \emph{Nature}, vol. 574, no. 7776, pp. 19--21, Oct. 2019.

\bibitem{Marinescu2005:Promise}
D.~C. Marinescu, ``The promise of quantum computing and quantum information theory - quantum parallelism,'' in \emph{Proc. IEEE IPDPS}, 2005, p. 112.

\bibitem{Grover1996:A_fast_quantum}
L.~K. {Grover}, ``{A fast quantum mechanical algorithm for database search},'' in \emph{Proc. ACM STOC}, 1996, pp. 212--219.

\bibitem{Alves2019:Quantum}
C.~Alves, L.~P. Santos, and T.~Bashford-Rogers, ``A quantum algorithm for ray casting using an orthographic camera,'' in \emph{Proc. ICGI}, 2019, pp. 56--63.

\bibitem{Haines2019:Ray}
E.~Haines and T.~Akenine-M{\"o}ller, \emph{Ray Tracing Gems: High-Quality and Real-Time Rendering with DXR and Other APIs}.\hskip 1em plus 0.5em minus 0.4em\relax Apress, 2019.

\bibitem{Glassner2001:Andrew}
A.~Glassner, ``Quantum computing, part 3,'' \emph{IEEE Comput. Graphics Appl.}, vol.~21, no.~6, pp. 72--82, 2001.

\bibitem{Kim23:Machine}
S.~Kim and J.~Seo, ``Machine-learning-based classification of {GPS} signal reception conditions using a dual-polarized antenna in urban areas,'' in \emph{Proc. IEEE/ION PLANS}, Apr. 2023, pp. 113--118.

\bibitem{Kim22:Machine}
S.~Kim, J.~Byun, and K.~Park, ``Machine learning-based {GPS} multipath detection method using dual antennas,'' in \emph{Proc. ASCC}, May 2022, pp. 691--695.

\bibitem{Lee22:Urban}
H.~Lee, J.~Seo, and Z.~Kassas, ``Urban road safety prediction: A satellite navigation perspective,'' \emph{IEEE Intell. Transp. Syst. Mag.}, vol.~14, no.~6, pp. 94--106, Nov.-Dec. 2022.

\bibitem{Kim23:Single}
S.~Kim, S.~Park, and J.~Seo, ``Single antenna based {GPS} signal reception condition classification using machine learning approaches,'' \emph{J. Position. Navig. Timing}, vol.~12, no.~2, pp. 149--155, 2023.

\bibitem{Lee23:Seamless}
Y.~Lee, Y.~Hwang, J.~Y. Ahn, J.~Seo, and B.~Park, ``Seamless accurate positioning in deep urban area based on mode switching between {DGNSS} and multipath mitigation positioning,'' \emph{IEEE Trans. Intell. Transp. Syst.}, vol.~24, no.~6, pp. 5856--5870, Jun. 2023.

\bibitem{Lee22:Optimal}
H.~Lee, S.~Pullen, J.~Lee, B.~Park, M.~Yoon, and J.~Seo, ``Optimal parameter inflation to enhance the availability of single-frequency {GBAS} for intelligent air transportation,'' \emph{IEEE Trans. Intell. Transp. Syst.}, vol.~23, no.~10, pp. 17\,801--17\,808, Oct. 2022.

\bibitem{Sun21:Markov}
A.~K. Sun, H.~Chang, S.~Pullen, H.~Kil, J.~Seo, Y.~J. Morton, and J.~Lee, ``Markov chain-based stochastic modeling of deep signal fading: Availability assessment of dual-frequency {GNSS}-based aviation under ionospheric scintillation,'' \emph{Space Weather}, vol.~19, no.~9, pp. 1--19, Sep. 2021.

\bibitem{Kim21:GPS}
S.~Kim, H.~Lee, and K.~Park, ``{GPS} multipath detection based on carrier-to-noise-density ratio measurements from a dual-polarized antenna,'' in \emph{Proc. ICCAS}, 2021, pp. 1099--1103.

\bibitem{Jia21:Ground}
M.~Jia, H.~Lee, J.~Khalife, Z.~M. Kassas, and J.~Seo, ``Ground vehicle navigation integrity monitoring for multi-constellation {GNSS} fused with cellular signals of opportunity,'' in \emph{Proc. IEEE ITSC}, 2021, pp. 3978--3983.

\bibitem{Park21:Single}
K.~Park and J.~Seo, ``Single-antenna-based {GPS} antijamming method exploiting polarization diversity,'' \emph{IEEE Trans. Aerosp. Electron. Syst.}, vol.~57, no.~2, pp. 919--934, Apr. 2021.

\bibitem{Lee24:A}
H.~Lee, S.~Kim, J.~Park, S.~Jeong, S.~Park, J.~Yu, H.~Choi, and J.~Seo, ``A survey on new parameters of {GPS CNAV/CNAV-2} and their roles,'' \emph{J. Position. Navig. Timing}, vol.~13, no.~1, pp. 45--52, 2024.

\bibitem{Santos2022:Towards}
L.~P. Santos, T.~Bashford-Rogers, J.~Barbosa, and P.~Navratil, ``Towards quantum ray tracing,'' \emph{IEEE Trans. Visual Comput. Graphics}, 2024, in press.

\bibitem{Nielsen2010:Quantum}
M.~L. Nielsen and I.~L. Chuang, \emph{Quantum Computation and Quantum Information}, 10th~ed.\hskip 1em plus 0.5em minus 0.4em\relax Cambridge University Press, 2010.

\bibitem{Shaik20:Implementation}
E.~h. Shaik and N.~Rangaswamy, ``Implementation of quantum gates based logic circuits using {IBM Qiskit},'' in \emph{Proc. ICCCS}, 2020, pp. 1--6.

\bibitem{Barenco1995:Elementary}
A.~Barenco, C.~H. Bennett, R.~Cleve, D.~P. DiVincenzo, N.~Margolus, P.~Shor, T.~Sleator, J.~Smolin, and H.~Weinfurter, ``{Elementary gates for quantum computation},'' \emph{Phys. Rev. A}, vol.~52, p. 3457, 1995.

\bibitem{McCluskey1956:Minimization}
E.~J. McCluskey, ``Minimization of {Boolean} functions,'' \emph{Bell Labs Tech. J.}, vol.~35, no.~6, pp. 1417--1444, 1956.

\bibitem{Petrick1956:Direct}
S.~R. Petrick, ``A direct determination of the irredundant forms of a {Boolean} function from the set of prime implicants,'' USAF Cambridge Research Center, Bedford, Mass., AFCRC Technical Report TR-56-100, April 1956.

\bibitem{Chandra1978:Number}
A.~K. Chandra and G.~Markowsky, ``On the number of prime implicants,'' \emph{Discrete Math.}, vol.~24, no.~1, pp. 7--11, 1978.

\bibitem{Rudell1987:Multiple}
R.~Rudell and A.~Sangiovanni-Vincentelli, ``Multiple-valued minimization for {PLA} optimization,'' \emph{IEEE Trans. Comput. Aided Des. Integr. Circuits Syst.}, vol.~6, no.~5, pp. 727--750, 1987.

\end{thebibliography}

\vspace{12pt}

\end{document}